\documentclass[journal,transmag]{IEEEtran}

\usepackage{microtype}

\ifCLASSINFOpdf
  \usepackage[pdftex]{graphicx}
  \graphicspath{{./images/}}
\else
\fi

\usepackage{svg}

\usepackage{amsmath,amssymb,amsfonts}

\usepackage{cleveref, stfloats}

\usepackage{tabularx,booktabs}
\newcolumntype{C}{>{\centering\arraybackslash}X}

\begin{document}

\title{Real-Time Detection of Drowsiness Among Vehicle Drivers: A Machine Learning Algorithm for Embedded Systems}

\author{\IEEEauthorblockN
  {Ashwin Pillay\IEEEauthorrefmark{1}\IEEEauthorrefmark{2},
    Aditya Kale\IEEEauthorrefmark{1}\IEEEauthorrefmark{3},
    Raj Anchan\IEEEauthorrefmark{1}\IEEEauthorrefmark{4},
    Aniket Bhadricha\IEEEauthorrefmark{1}\IEEEauthorrefmark{5},
    Sangeeta Prasanna Ram\IEEEauthorrefmark{1}\IEEEauthorrefmark{6}}
  \IEEEauthorblockA{\IEEEauthorrefmark{1}Dept. of Instrumentation, Vivekanand Education Society’s Institute of Technology}
  \IEEEauthorblockA{\IEEEauthorrefmark{2}2016.ashwin.pillay@ves.ac.in}
  \IEEEauthorblockA{\IEEEauthorrefmark{3}2016.aditya.kale@ves.ac.in}
  \IEEEauthorblockA{\IEEEauthorrefmark{4}2016.raj.anchan@ves.ac.in}
  \IEEEauthorblockA{\IEEEauthorrefmark{5}2016.aniket.bhadricha@ves.ac.in}
  \IEEEauthorblockA{\IEEEauthorrefmark{6}sangeeta.prasannaram@ves.ac.in}
  }

\IEEEtitleabstractindextext{%
\begin{abstract}
Numerous studies have established the necessity for developing safety equipment to detect drowsiness among vehicle drivers. However, for reliable implementations, such systems must employ dependable sources of stimuli— through Electrooculography (EOG), the tendencies of drowsiness can be directly sensed by measuring blinks of prolonged durations. While conventional machine learning (ML) algorithms can be utilized for the detection and classification of these prolonged blinks (PB), executing them on microcontroller units (MCU) may prove to be a laborious task. Hence, by keeping resource constraints and practicality in mind, an ML algorithm is proposed in this study to identify PBs executed by an individual with desirable accuracy and precision while being efficient enough to be deployed on portable wearables using economic MCUs. Furthermore, the suggested algorithm is subjected to multiple rounds of testing in this study thereby, establishing its possibility as a feasible drowsiness detection measure for wearable systems.
\end{abstract}

}

\maketitle

\IEEEdisplaynontitleabstractindextext

\IEEEpeerreviewmaketitle

\section{Introduction}

\IEEEPARstart{T}{he} eyes are among the most vital organs in the human body, aiding as
a sensing element in carrying out a plethora of quotidian tasks,
including spatial awareness and vehicular driving. More remarkably, the
eyes are as much an indicator of the human state as it is a sensor:
the ocular activity of a person may indicate numerous different
parameters such as their alertness, expressions, indications, mental
and physical conditions. To this end, several technological
applications have been developed based on the analysis of various eye
movements and actions \cite{1}\cite{2}\cite{3}.

In fact, gauging the alertness of a subject based on their eye
activity is among the most elementary tasks performed by
humanity. This procedure can be extended to situations where
identifying a subject’s drowsiness is indispensable, like driving
vehicles or operating heavy machinery.  Existing literature has
successfully identified a high correlation between drowsiness and
blinks of prolonged durations, referred to as Prolonged Blinks (PBs)
in this study \cite{4}. By detecting about two to three such PBs in a ten to
fifteen-second interval using eye-activity measurement techniques such
as electrooculography (EOG) \cite{5}, the probability of a driver being
drowsy may be ascertained. Consequently, they can be alerted before an
impending mishap.

However, vehicular driving is an activity where the eye is employed
for multiple services, ranging from achieving spatial awareness in the
environment to gauging various instruments on the vehicle to control
it as desired. EOG will naturally account for all these eye movements;
hence, isolating only those electrooculograms that correspond to PBs
becomes essential for reliable detections of drowsiness. Furthermore,
for dependability and affordability, such isolation and ensuing
detection procedures should be carried out by portable, isolated
embedded systems. However, they are generally disadvantaged in both
data processing power and storage.

With these considerations, an efficient MCU algorithm for the
isolation of EOG waveforms corresponding to PBs and onward detection
of drowsiness is discussed in this paper, which is organised as follows:
\cref{sec:setup} describes the instrument setup for detection of PBs. \Cref{sec:isolation} introduces the initial preprocessing scheme employed on the EOG waves and expounds on the distinct properties of PB EOG waveforms used by the algorithm to differentiate them from other eye movements. \Cref{sec:isolation} briefs about the execution of the drowsiness detection algorithm, emphasising on its machine-learning (ML) phase. \Cref{sec:optimization} presents the results of executing this algorithm both as a proof of concept and as a practical implementation. Finally, \cref{sec:results} culminates the study by submitting the conclusions and future scope of the algorithm based on the observed results.

\section{Setup}
\label{sec:setup}

For this study, a three-electrode assembly (two measuring electrodes
and one reference electrode) was fastened onto the optimal positions
of the subject's face to detect PBs \cite{6}. These electrodes were then
interfaced with a custom made signal conditioning PCB that selectively
amplifies signals lying in the EOG spectrum while filtering out all
other unwanted waveforms. The processed output was then provided as
input to a 12-bit microcontroller unit (MCU) to carry out the PB
isolation and detection operations.

\section{Isolation of Desired EOG Waveforms}
\label{sec:isolation}

\subsection{Pre-Processing}
\label{sec:preprocessing}

EOG waveforms initially received by the MCU $x$ are prone to baseline
wanders similar to other biosignals like ECG \cite{7}. Hence, any analysis
based on the grounds of their amplitudes alone is
ineffective. Moreover, these waves are also prone to be superimposed
by the noise produced from sources like motion artefacts, EMI and
improper attachment of electrodes on the face. In any case, reliable
detection of PBs depends more on the tangible properties of the EOG
wave received than on the subtler details. Hence, there is an
opportunity to improve detection results by allowing some
pre-processing to be performed on the incoming signal before analyzing
it further.

For the ease in analysis and optimal utilization of MCU resources,
therefore, the EOG signal is initially processed by a modified, moving
average filter \(M\{x(n)\}\) of window size \(N = 25\) such that for every
sample of the original signal, \(x(n)\), the filter results in a smoothed
equivalent \(x_{avg}(n)\):

\begin{IEEEeqnarray}{rCl"s}
  \label{eq:1}
  x_{\textnormal{avg}}(n)&=& M\{x(n)\}\qquad\qquad
  \\
  &=&\left\{ \,
    \begin{IEEEeqnarraybox}[][c]{l?s}
      \IEEEstrut
      x(n); & if $|x(n)-\>\bar x_N(n)| > r\_thresh$ \nonumber\\
      \bar x(n); &  $otherwise$ 
      \IEEEstrut

    \end{IEEEeqnarraybox}
  \right.
  \label{eq:Moving avg filter}
\end{IEEEeqnarray}
\textrm{where;}
\begin{IEEEeqnarray}{rCl}
  x(n)&=&$n\textsuperscript{th} sample of EOG signal$\nonumber\\
  r\_thresh&=&$regularization threshold; constant that$\nonumber\\&& $controls degree of smoothing obtained$\nonumber \\
  \bar x_N(n)&=& $mean of previous N\textsuperscript{th} samples; stored$\nonumber\\&& $ in a buffer size of N$\nonumber\\
  &=& \frac{1}{N} \sum_{i=n-N}^{n-1} x(i)\nonumber\\
  N &=& $min(n,25)$\nonumber\\
  x_{\textnormal{avg}}(n)&=& $filtered (smoothed) equivalent of x(n)$\nonumber
\end{IEEEeqnarray}

In equation \ref{eq:1}, \(r\_thresh\) denotes the regularization
threshold—a positive, floating-point valued and controllable parameter
determining the extent of smoothing achieved by the filter. As
\(r\_thresh\) is increased upwards from zero, the degree of smoothing
performed on \(x_{avg}\) will increase. For applications requiring
subtler details in the smoothed waveform, a low value of \(r\_thresh\)
must be set. After a round of experimental trials, \(r\_thresh\) was
suitably set to 1 for this study.

Once the wave is smoothed, its first-order derivative (FOD) is
obtained to remove the complications posed by the wandering baseline.

\begin{IEEEeqnarray}{rCl"l}
  \label{eq:2}
  x_{\textnormal{FOD}}(n)&=& F\{x_{\textnormal{avg}}(n)\}\qquad\qquad\\
  &=&\left\{ \,
    \begin{IEEEeqnarraybox}[][c]{l?s}
      \IEEEstrut
      x_{\textnormal{avg}}(n) -\> x_{\textnormal{avg}}(n-N); & $x_{\textnormal{avg}}(n)$\\&$ -\> x_{\textnormal{avg}}(n-N)$\\
      & $>FOD$\\&$\_clearance$\nonumber\\
      &$\_threshold$ \\
      0; &  $otherwise$

      \IEEEstrut

    \end{IEEEeqnarraybox}
  \right.
\end{IEEEeqnarray}
\textrm{where;}
\begin{IEEEeqnarray}{rCl}
  FOD\_clearance\_threshold&=&$clearance threshold that $\nonumber\\&&$\nonumber controls sensitivity parameter $\\&&$\nonumber for calculation of FOD;$\\&&$\nonumber if FOD\_clearance\_threshold $\\&&$\nonumber is low, higher will be$\\&&$\nonumber the sensitivity$\nonumber\\
  N &=& $min(n,25)$\nonumber\\
  x_{\textnormal{FOD}}(n)&=& $FOD of filtered $ x_{\textnormal{avg}}(n)$ sample $\nonumber\\&&$for convenience; $\nonumber
\end{IEEEeqnarray}

The adjustable, positive valued \(FOD\_clearance\_threshold\) in
equation \ref{eq:2} sets the sensitivity of \(F(n)\). Higher values
of this threshold will cause only the more sudden and significant
changes of the EOG signal to appear in $x_{FOD}$. In
comparison, smaller values (closer to zero) will also consider the
more gradual deviations in the output signal. After a round of
experimental trials, the \(FOD\_clearance\_threshold\) was set to 0.1
for this study.

For the convenience of this research, consider equation \ref{eq:3}
\begin{IEEEeqnarray}{rCl}
  \label{eq:3}
  \mathrm{let}\ r(n) = x_{FOD}(n)
\end{IEEEeqnarray}

The signal \(r\) is the resultant obtained from the pre-processing stage;
all ensuing analysis will be performed on \(r\).

\begin{figure}
  \centering
  \includegraphics{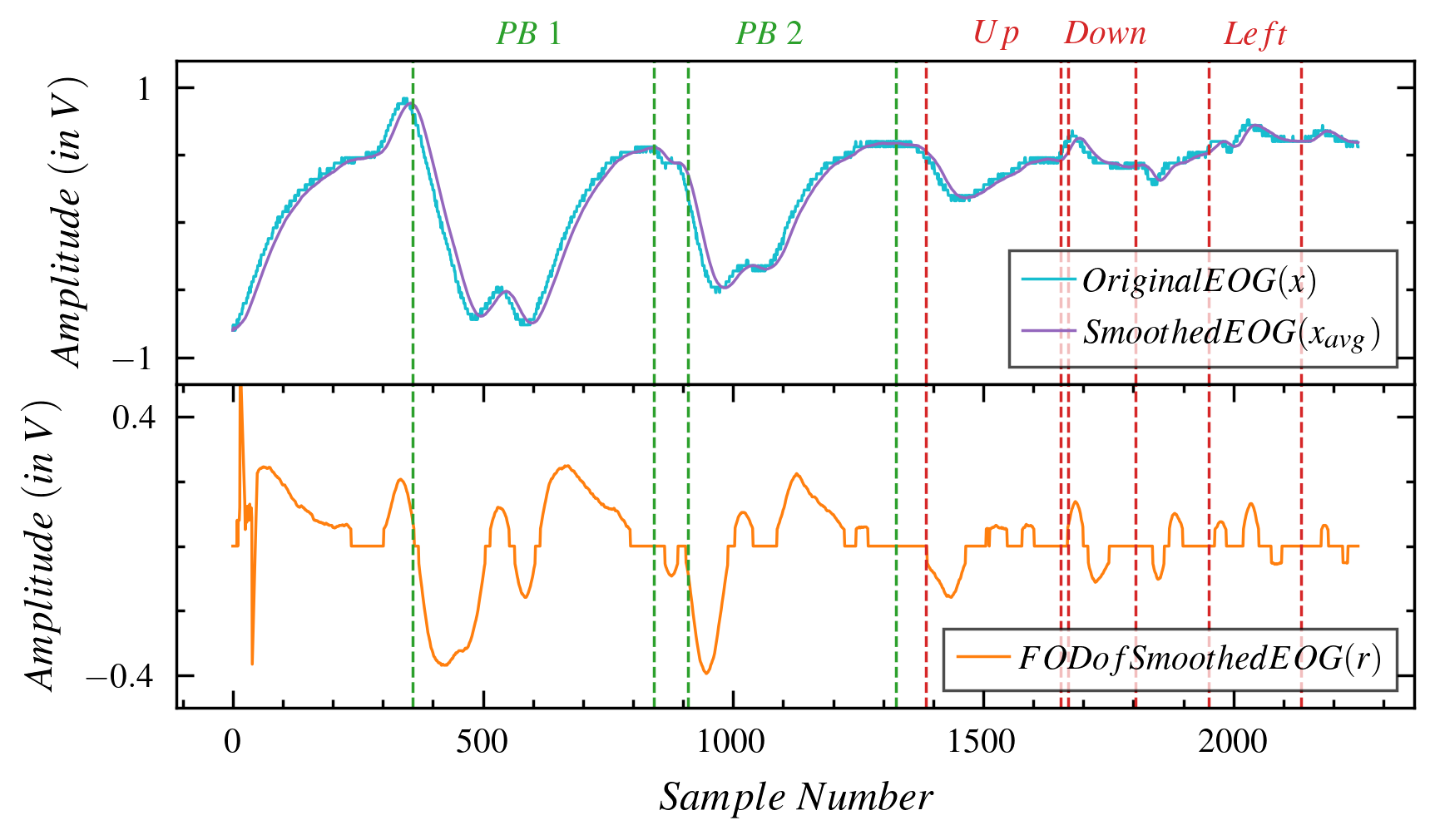}
  \caption{Original EOG Signal $x$ and Resulting FOD Signal $r$ after Pre-Processing}
  \label{fig:1}
\end{figure}

\subsection{Understanding Idiosyncratic Characteristics of PBs}
\label{sec:idiosyncracies}

\begin{figure}
  \centering
  \includegraphics{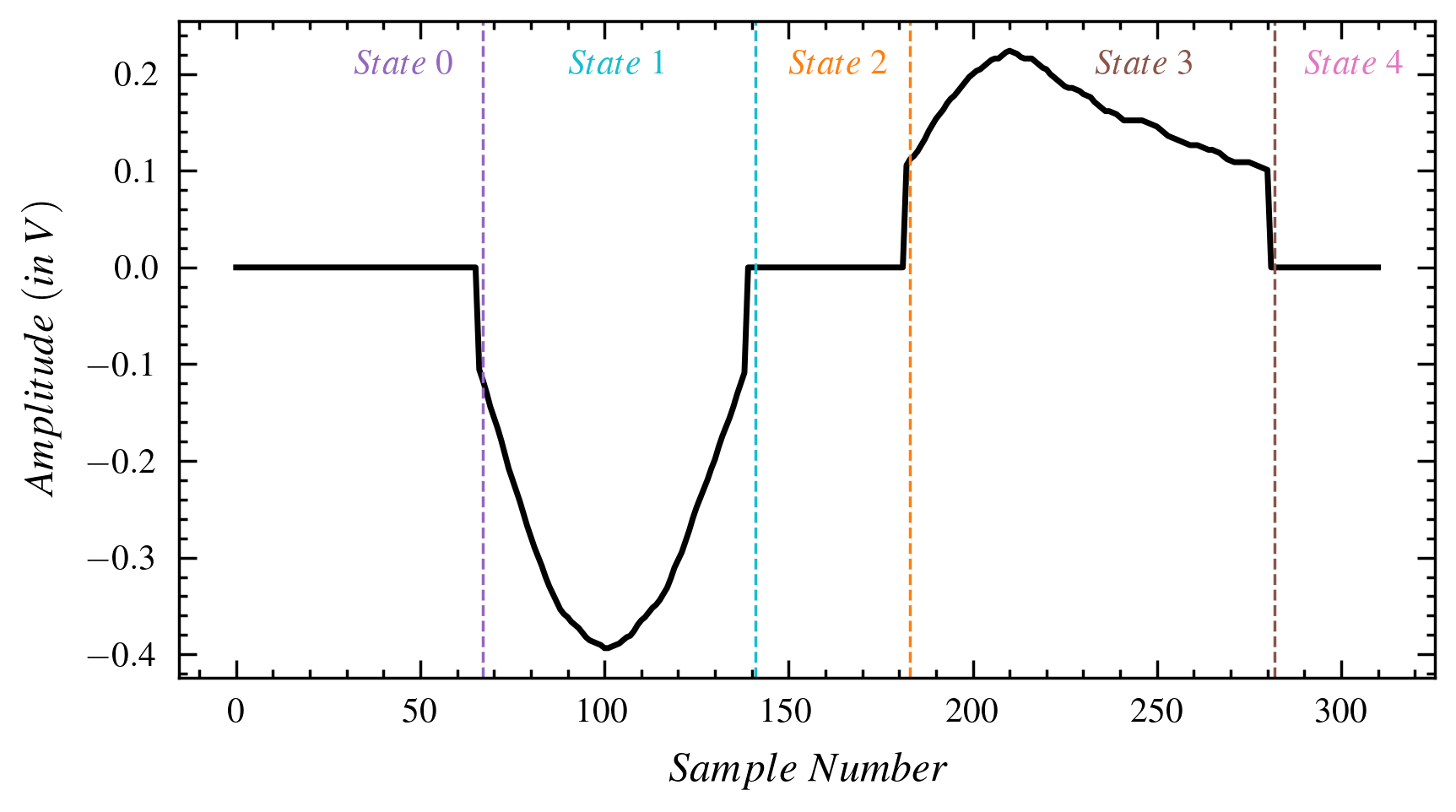}
  \caption{FOD of a Prolonged Blink EOG signal and its State Definition}
  \label{fig:2}
\end{figure}

To isolate exclusively those sections of the EOG FOD signal \(r\) that
correspond to PBs, the unique traits of the same are identified and
defined— this is done by defining a set of states that the PB wave is
observed to strictly pass through.  Initially, when the subject
executes no eye movements, the signal amplitude is on the zero-line or
the reference line and is designated to be in State 0. Then, as
observed in \cref{fig:2}, the FOD of PBs is characterized by an initial
negative excursion that rises back to the zero line after reaching the
minima. This entire negative half cycle (NHC) of the PB is defined as
State 1. After traversing the NHC, there is a short period where the
value of \(r\) is zero; this region is labelled as the inter-half cycle
(IHC) zone and is represented by State 2. Corresponding to the NHC,
the PB wave also voyages through a positive half cycle (PHC) where the
amplitude reaches the wave maxima; the entirety of this PHC is
demarcated to be State 3. Finally, the FOD again drops back to the
zero line after State 3, assuming the final State 4.

Since all PBs generated by the subject are bound to cause the FOD signal to pass from States 0-4 sequentially, it is a trait that can be used to distinguish and isolate PBs from all other eye movements, except for the vigorously exhibited upward gazes. An algorithm that ensures that all PB candidate waveforms pass through these defined states may, therefore, act as an initial filter to prevent non-PB eye movements from being analyzed further. This facilitates the optimum use of MCU resources and improves overall detection efficiency. The algorithm carries out the State evaluation procedure till the wave remains in State 4 continuously for 200 ms, rejecting non-candidate waves as soon as they disobey the sequential transition scheme.

\section{Classification of EOG Waveforms}
\label{sec:classification}

\subsection{Initial Subject-specific Learning}
\label{sec:learning}

Like most physiological signals \cite{8}, the characteristics of EOG signals
also vary from one individual to another. Such variations are the
result of multiple factors, including differences in muscle strength,
age, fatigue or stress level, gender \cite{9} \cite{10} \cite{11}. As a result, while
PBs generated by different subjects have the same skeleton—as
described by the four States defined previously, their subtleties may
differ. Any outright classification scheme that does not account for
these distinctions may hence, not yield the reliability expected in
practical applications; for example, misclassifying eye movements with
similar EOG waveforms like the upward gaze and PB.

Consequently, to understand and analyze subject-specific signals, the
classification algorithm must employ artificial intelligence
(AI). However, for an isolated embedded system, executing conventional
machine learning (ML) algorithms like multilayer artificial neural
networks (ANNs) may pose to be a strenuous task. Therefore, to suit
the portable detection systems in focus of this study, a novel
algorithm is developed. Firstly, it learns the distinct features of
EOG signals generated by a given user and subsequently employs this
knowledge for successfully classifying PBs in real-time without
exhausting the available resources at any point during its execution.

The learning scheme employed is inspired by the widely established and
efficient strategy used to train fingerprint detectors in
smartphones \cite{13}. Initially, \(wavelet\_buffer\), a matrix that will act as
a record of most identical PBs, is defined in the MCU memory. As a
first step, the user will be expected to execute a set number
\(prolonged\_training\_reps\) of PBs. In this period \(lperiod\_1\), the
algorithm implemented by the MCU will assume all eye movements
generated and undergoing a sequential transition from State 0-4 to be
PBs; the samples of the corresponding EOG wave being stored in an
array wavelet. If \(wavelet\_buffer\) is empty, wavelet is added to it
without further analysis; however, if it is populated, the
\(most\_similar\_wave\) among its members is compared with wavelet: the
similarity or correlation coefficient between them being calculated
and assigned to \(similarity\_measure\). The \(most\_similar\_wave\) is obtained
conveniently by determining the medoid wave of wavelet\_buffer \cite{14},
while the comparison is quantified by DSP algorithms like maxima of
normalised cross-correlation \cite{15} or normalised DDTW with Sakoe-Chiba
based constraint \cite{10} \cite{17} \cite{18} depending on the suitable computational
complexity versus accuracy compromise \cite{19}.

The correlation coefficient is only one among six features
(\(total\_features\) = 6) analysed in wave $r$ for this study; these are
outlined in \cref{tab:t1}. Depending on the strictness needed for
the classification of waves, the number of features analysed can be
varied as required. For each feature during \(lperiod\_1\), the mean
and standard deviation (SD) parameters are updated as per equations 4
and 5. The updated values of these statistics are subsequently
maintained in their respective buffers.

\begin{table*}[!ht] \setlength\tabcolsep{0 pt} 
  \caption{The set of features in EOG wave $r$ that are analyzed for detecting PBs}
  \label{tab:t1}
  \begin{center}
    \begin{tabularx} {\textwidth}{l*{4}{|C}}
      \hline
      & Feature & Associated Variable & PB Characteristic Represented [4] \\
      \hline
      & Similarity/Correlation Coefficient & \(most\_similar\_wave\) & Skeleton of PB \\
      \hline
      & Positive Peak of \(r\) & \(max\) & Percentage of Eye Closure \\
      \hline
      & Negative Peak of \(r\) & \(min\) & Percentage of Eye Closure \\
      \hline
      & Positive Half Cycle (PHC) Duration of \(r\) & \(p\_durn\) & Eyelid closing speed \\
      \hline
      & Negative Half Cycle (NHC) Duration of \(r\) & \(n\_durn\) & Eyelid reopening speed \\
      \hline
      & Total Wave Duration of \(r\) & \(t\_durn\) & Total PB duration \\
      \hline
    \end{tabularx}
  \end{center}
\end{table*}

For each feature,
\begin{IEEEeqnarray}{rCl}
  \label{eq:4}
  $mean (N)$ &=& \frac{mean(N-\>1)\times(N-\>1)+\>value(N)}{N}\enskip\qquad
  \label{eq:Mean Calc}\\ 
  $where;$\nonumber\\\nonumber
  $N$&=& $current reading number$
\end{IEEEeqnarray}
\\
For each feature,
\begin{IEEEeqnarray}{rCl"s}
  \label{eq:5}
  variance (N) &=& variance (N-1)+\>(value(N)-\>mean(N))^2\nonumber\\
  SD (N) &=&\frac{variance(N)}{N}
\end{IEEEeqnarray}

Finally, \(wavelet\) is appended as a new entry of the \(wavelet\_buffer\), and
the wave counter \(total\_readings\) is incremented before analyzing the
next eye movement.

Following \(lperiod\_1\), the user must generate a predefined number
\(up\_training\_reps\) of upward gazes. In this phase
(\(lperiod\_2\)), the same set of features are analyzed in wave
r. Here, the statistical parameters \(anti\_mean\) and \(anti\_SD\)
for each feature are updated similar to mean and variance of PBs,
respectively (referring to equations \ref{eq:4} and \ref{eq:5}). While
the analysis of PBs in \(lperiod\_1\) is used to set the thresholds for
their accurate real-time classification, the data acquired in
\(lperiod\_2\) is used to identify the idiosyncrasies of up movements and
thereby make a more comprehensive differentiation between them and PB
waveforms. After a round of experimental trials, the values of both
\(PB\_training\_reps\) and \(up\_training\_reps\) were conveniently set to ten.

Together \(lperiod\_1\) and \(lperiod\_2\) constitute the $Learning
\ Period$ of the algorithm, which should last for around one-two minutes
depending upon the subject's consistency. Post this period, the
thresholds for classifying PBs from other eye movements can be
determined. For each feature examined: the upper (ut) and lower
thresholds (lt) and the upper (uat) and lower (lat) anti-thresholds
are calculated as per equations 6a and 6b using the obtained mean,
variance, \(anti\_mean\) and \(anti\_SD\). This threshold adjusting
scheme employed will also account for situations where the threshold
bands of PB and up waves of a feature merge, assigning the midpoint of
the merging region as the corresponding threshold for that PB feature,
as proposed by equations 7a and 7b.

\begin{IEEEeqnarray}{rCl"s}
  lt(N)&=&mean(N)-\>(1.5\times S.D(N))\nonumber\\
  ut(N)&=&mean(N)+\>(1.5\times S.D(N))\IEEEyesnumber \IEEEyessubnumber\\
  \label{eq:thresh}\nonumber\\
  lat(N)&=&anti\_mean(N)-\>(1.5\times anti\_S.D(N))\nonumber\\
  uat(N)&=&anti\_mean(N)+\>(1.5\times anti\_S.D(N)) \IEEEyessubnumber\\ \nonumber\\\nonumber
\end{IEEEeqnarray}

\begin{IEEEeqnarray}{rCl"s}
  \label{eq:7a}
  lt(N)&=&\left\{ \,
    \begin{IEEEeqnarraybox}[][c]{l?s}
      \IEEEstrut
      \frac{lt(N)+\>uat(N)}{2};\nonumber\\
      \qquad ut(N) >\> uat(N) >\>
      \IEEEyesnumber \IEEEyessubnumber\\ 
      \qquad lt(N) >\> lat(N)
      \\
      lt(N);\qquad $otherwise$ 
      \IEEEstrut
    \end{IEEEeqnarraybox}
  \right.
\end{IEEEeqnarray}
\begin{IEEEeqnarray}{rCl"s}
  \label{eq:7b}
  ut(N)&=&\left\{ \,
    \begin{IEEEeqnarraybox}[][c]{l?s}
      \IEEEstrut
      \frac{ut(N)+\>lat(N)}{2};\nonumber\\
      \qquad uat(N) >\> ut(N) >\>
      \IEEEyessubnumber\\ 
      \qquad lat(N) >\> lt(N)
      \\
      ut(N);\qquad $otherwise$ 
      \IEEEstrut
    \end{IEEEeqnarraybox}
  \right.
\end{IEEEeqnarray}

Once the $Learning\ Period$ is completed and thresholds are set, the
system is ready to detect the occurrence of ensuing PBs in real-time.

\subsection{Real-time Detection ofPBs}
\label{sec:detection}

\begin{figure}[b]
  \centering
  \includegraphics{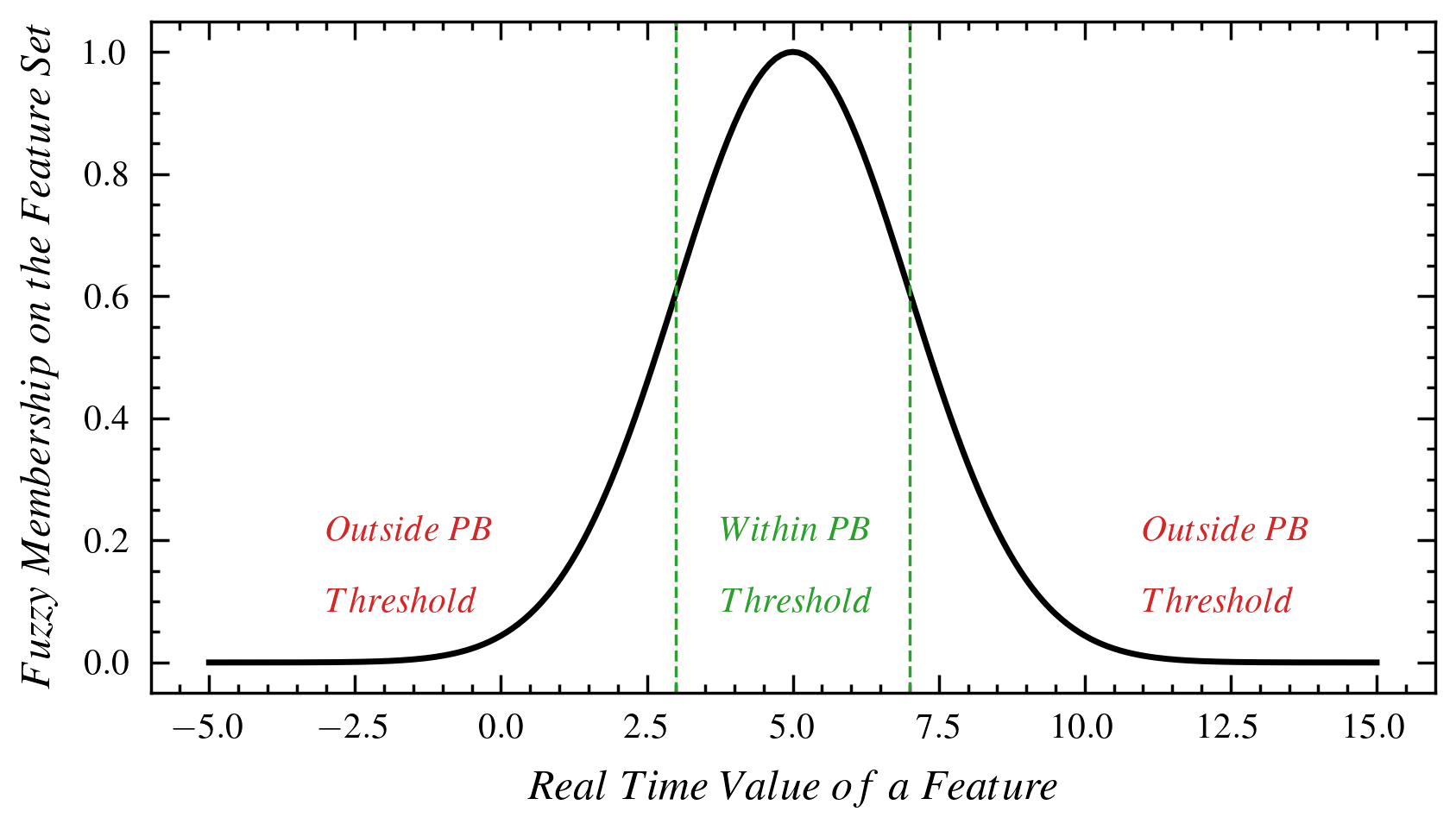}
  \caption{Gaussian function used for determining the membership of a candidate EOG wave’s feature on its corresponding PB set}
  \label{fig:3}
\end{figure}

The thresholds \(lt\) and \(ut\) set in the $Learning\ Period$ are used to
classify any PBs generated by the same subject in real-time, referred
to as the $Operational\ Period$. For any generated EOG wave to be
considered as a PB, it should initially pass from State 0 - 4,
followed by having all the aforementioned features within tolerable
limits set by the corresponding thresholds for that
feature. Practically it is observed that true PBs generated by the
user may, in some instances, have some of the considered attributes
outside their thresholds. To avoid such PBs from being
misclassified, fuzzy logic is employed over crisp logic, ensuring that
a holistic classification is made on the closeness of all the features
of the EOG wave with the existing dataset.

For an EOG wave obtained during the $Operational\ Period$, the membership
degree of each feature on its corresponding PB set is obtained using a
gaussian membership function given by equation 8. The membership
degree of all the features considered is then added up and assigned to
\(pass\_sum\). The defuzzification necessary for the classification of PB
is finally performed by equation 9, resulting in a binary output for
the entire analysis that answers whether the eye movement generated
was a PB or not.

For each feature,
\begin{IEEEeqnarray}{rCl"s}
  \label{eq:8}
  \nonumber\\
  fuzz\_val(N) = exp(-\frac{1}{2} \cdot (\frac{value(N)-\frac{ut(N)+lt(N)}{2}}{\frac{ut(N)-lt(N)}{2}}))
\end{IEEEeqnarray}

\begin{IEEEeqnarray}{rCl"s}
  \label{eq:9}
  is\_PB(N)&=&\left\{ \,
    \begin{IEEEeqnarraybox}[][c]{l?s}
      \IEEEstrut
      TRUE;\qquad \frac{pass\_sum}{total\_features} \geq 0.6\nonumber\\
      FALSE;\qquad  $otherwise$ \IEEEyesnumber
      \IEEEstrut
    \end{IEEEeqnarraybox}
  \right.
\end{IEEEeqnarray}

\begin{table*}[bp] \setlength\tabcolsep{0 pt}
  \caption{Algorithm Test Results for Different Subject Profiles}
  \label{t2}
  \begin{center}
    \begin{tabularx} {\textwidth}{l*{9}{|C}}
      \hline
      & Subject Profile & Total Readings & Correct Detections & Wrong Detections & False Positives & True Negatives & Avg. Time per Detection (ms) & \% Accuracy \\ 
      \hline
      & A & 398 & 310 & 78 & 4 & 74 & 441 & 79.90 \\
      \hline
      & B & 383 & 343 & 30 & 17 & 13 & 428 & 91.96 \\
      \hline
      & C & 320 & 282 & 28 & 18 & 10 & 443 & 90.97 \\
      \hline
      & D & 321 & 290 & 21 & 17 & 4 & 416 & 93.25 \\
      \hline
      & E & 325 & 263 & 52 & 0 & 52 & 499 & 83.49 \\
      \hline
      & F & 302 & 260 & 32 & 31 & 1 & 389 & 89.04 \\
      \hline
      & G & 328 & 273 & 45 & 14 & 31 & 369 & 85.85 \\
      \hline
      & H & 383 & 329 & 44 & 0 & 44 & 374 & 88.20 \\
      \hline
      & I & 364 & 306 & 48 & 41 & 7 & 445 & 86.44 \\
      \hline
      & J & 398 & 264 & 124 & 0 & 124 & 380 & 68.04 \\
      \hline
      & K & 313 & 262 & 41 & 23 & 18 & 482 & 86.47 \\
      \hline
      & L & 326 & 285 & 31 & 11 & 20 & 382 & 90.19 \\
      \hline
      & M & 398 & 359 & 29 & 6 & 23 & 430 & 92.53 \\
      \hline
      & N & 317 & 261 & 46 & 37 & 9 & 437 & 85.02 \\
      \hline
      & O & 302 & 246 & 46 & 34 & 12 & 425 & 84.25 \\
      \hline
    \end{tabularx}
  \end{center}
\end{table*}

\section{Optimizing the Detection Algorithm for MCUs}
\label{sec:optimization}

The algorithm described in sections 3 and 4 was executed by the TIVA
TM4C123G, a 32-bit ARM® Cortex® M4 MCU operating at 80MHz, coupled
with a 32kB SRAM and 2kB EEPROM \cite{20}. Considering the EOG wave samples
to be of two-byte short datatype and each EOG wave to amass about 500
samples on an average, the MCU (executing the unoptimized program) was
observed to run out of available RAM after recording just 8-15
readings. Such performances are unsuitable since the accuracy in
determining the medoid wave (for Similarity Coefficient calculation)
is sufficient only when enough PB waves are maintained in the
buffer. Moreover, this calculation occurs for every new reading, and
so, the MCU must always keep this buffer in memory for quick detection
of drowsiness.

Memory fragmentation is another issue that needs to be addressed to
ensure the long term viability of the MCU algorithm. Considering the
limited RAM, discontinuous data structures like linked lists are not
practical due to the 4 bytes utilized per pointer. For continuous
arrays, creating space for new waves by freeing the memory occupied by
another wave will be optimized only when such waves have the same
length. This is a rare occurrence since real-time PBs can have any
random number of samples per wave. Fragmentation may also occur while
using moving buffers for mean filtering and FOD calculation.

One of the solutions identified for the aforementioned problems was to
optimize the datatype used for each variable. For example, most
variables were associated with the short datatype that occupies two
bytes instead of the default int datatype occupying four bytes.  Apart
from minimizing the number of float calculations part of the
algorithm, All data requiring continuously moving buffers are stored
in circular buffers that occupy a predefined range of
RAM. Furthermore, as a compromise between linked lists and plain
arrays, a hashed array tree (HAT) with leaves of hundred elements were
used to store EOG waves. Using HATs with fixed leaf size prevents
memory fragmentation while ensuring no more than 49 memory elements
are left utilized at any time.

Another crucial factor governing the suitability of a candidate MCU
program for time-critical applications like drowsiness detection and
alerting is its execution time. Hence, the various calculations
performed by the code were optimized to the greatest extent, involving
minimal copying of data among the sub-modules. For example, phase
calculations were omitted in the cross-correlation algorithm used for
EOG wave comparison. They were also rewritten to comprise integer
based calculations only. Such optimizations ensure that the user is
alerted as soon as the established criteria of drowsiness are
satisfied.

\section{Results}
\label{sec:results}

While laboratory evaluations of the algorithm yielded acceptable
results, physically testing the same on vehicle drivers became
difficult due to the COVID-19 related lockdowns in India.\footnote{Results will be updated with the data obtained from real-world testing once physical experimentation is feasible.} However, to demonstrate its efficacy, a simulator was used to initially train the algorithm with PBs and upward gazes and then randomly generate any common eye movement to evaluate its performance.

The simulated signals for each eye movement have more than ninety per
cent correlation with the actual EOG signals for that
movement. Additionally, it was programmed to make minor variations in
the simulated signals to resemble the fluctuations in vigour and speed
that a human subject is likely to exhibit while using the
wearable. For each eye movement that the simulator generated, the
algorithm is expected to classify it as a PB or not correctly. The
results obtained from this comparison were then recorded as one among
a correct classification, a false positive, a true negative or an
unclassified detection (when the algorithm did not detect the
simulated wave).

For demonstrating the algorithm's suitability for different kinds of subjects, the simulator was used to recreate the behaviour of fifteen different subjects. The algorithm was individually trained with ten PBs and ten upward gazes for each subject profile and then executed for more than 300 readings. To ensure that the results resembled the real conditions as far as possible, the entire simulation process and the detection algorithm was run on the TIVA TM4C123G MCU. The results of this experiment are summarized in \cref{t2}.

\section{Conclusion}
\label{sec:conclusion}

From the aforementioned observations, it can be concluded that the algorithm discussed in this study can detect PBs with an average accuracy of 86\%. Moreover, the experiments conducted were documented for detecting individual PBs. Since drowsiness as a phenomenon is associated with the repeated exhibition of PBs (about two to three such PBs in a ten to fifteen-second interval), the overall effect of true negatives is averaged out, further reducing their repercussions on reliable drowsiness detection.

Moreover, unlike PBs, the upward gaze is a far more likely eye movement that a vehicle driver can perform. This may be due to actions like the periodic glancing at the rear-view mirror or looking at the signal indicated by traffic lights. Hence, it is necessary to ensure that no upward gazes are classified incorrectly as PB (false positive); failing to do so might repeatedly trigger the safety mechanism of the wearable, which can affect the efficiency and concentration of the user while driving. From the results obtained, it can be seen that the probability of false positives being generated by the algorithm is low.  Hence, it is unlikely that the system falsely detects consecutive upwards gazes.

Additionally, one of the most crucial metrics that may be used to benchmark safety systems is the time it requires to detect a hazardous situation. In this case, the algorithm manages to make its decisions within half a second when executed on the TIVA TM4C123G. Hence, even while using an economical MCU, the driver is provided with sufficient time to recover from their drowsiness and prevent an impending mishap from happening.

While the detection algorithm constitutes multiple optimizations and tradeoffs for the calculation and storage of data on isolated embedded systems, it manages to incorporate ML for addressing subject-specific characteristics present in the EOG signals. Additionally, considering the high accuracy and low detection time required for detecting PBs, it can be concluded that the algorithm proposed in this study may be used as part of an economic wearable system to detect drowsiness among vehicle drivers.

\subsection{Future Scope}
\label{sec:scope}

There are some aspects where the algorithm could be modified to
enhance its performance. Over time, an individual may exhibit varying
physical and mental health conditions due to causes like fever,
fatigue, age, distraction and stress. In these conditions, the eye
movements exhibited by them may not be of the same vigour and speed as
they did during the $Learning\ Period$ \cite{8} \cite{9}. If the
algorithm can fine-tune the feature thresholds according to the
short-time EOG wave trends observed during these situations,
reliability in detecting PBs can be ensured irrespective of user
conditions. Theoretically, such algorithms would run indefinitely and
may be developed using statistical means similar to those discussed in
\cref{sec:learning} a but gathered over a short-time window.

Additionally, it must be noted that the algorithm discussed in this study assumes a single wearable unit to be used by a single user only (for whom it is trained). However, adding a user profiling system (availing auxiliary permanent memory storage for the MCU) allows multiple user data to be maintained simultaneously. If implemented, a single wearable can then be used interchangeably by numerous individuals simply by switching to their user profile through button-based menu toggling.

\appendices

\ifCLASSOPTIONcaptionsoff
  \newpage
\fi

\bibliographystyle{IEEEtran}
\bibliography{library.bib}

\end{document}